\documentclass{aa}
\usepackage{graphicx,natbib,psfig}
\usepackage{txfonts}
\newcommand{\ltsima} {$\; \buildrel < \over \sim \;$}
\newcommand{\gtsima} {$\; \buildrel > \over \sim \;$}
\newcommand{\lta} {\lower.5ex\hbox{\ltsima}}
\newcommand{\gta} {\lower.5ex\hbox{\gtsima}}

\newcommand{\kms}{$\rm{\,km \,s}^{-1}$}

\begin{document}

\title{Measuring supermassive black holes with gas kinematics:\\
the active S0 galaxy \object{NGC 3998}\thanks{Based  on observations obtained at
the  Space  Telescope Science  Institute, which is operated by the
Association of Universities for Research in Astronomy, Incorporated,
under NASA contract NAS 5-26555.}\fnmsep 
\thanks{This publication makes use of the 
HyperLeda database, available at http://leda.univ-lyon1.fr.}}

\author{Giovanna De Francesco
\inst{1}
\and
Alessandro Capetti
\inst{1}
\and
Alessandro Marconi \inst{2}}

\offprints{G. De Francesco}

\institute{INAF - Osservatorio Astronomico di Torino, Strada
  Osservatorio 20, I-10025 Pino Torinese, Italy\\
\email{defrancesco@to.astro.it,capetti@to.astro.it}
\and
INAF - Osservatorio Astrofisico di Arcetri
       Largo E. Fermi 5, I-50125 Firenze, Italy\\
\email{marconi@arcetri.astro.it}}

\date{Received / Accepted}

\abstract
{We present results from a kinematical study of the gas in the
nucleus of the active S0 galaxy NGC 3998 obtained from archival HST/STIS
long-slit spectra.} 
{We analyzed the emission lines profiles and derived
the map of the gas velocity field.  The observed velocity curves are
consistent with gas in regular rotation around the galaxy's center.}
{By modeling the surface brightness distribution and
rotation curve of the H$\alpha$ emission line we found that the
observed kinematics of the circumnuclear gas can be accurately
reproduced by adding to the stellar mass component a compact dark mass
(black hole) of $M_{\rm BH} = 2.7_{-2.0}^{+2.4}\times 10^8 
 M_{\odot}$ (uncertainties at a 2 $\sigma$ level); the radius of its sphere of
influence ($R_{\rm sph} \sim$ 0\farcs16) is well resolved at the HST resolution.} 
{The BH mass estimate in NGC 3998 is in good agreement with
both the $M_{\rm BH}-M_{\rm bul}$ (with an upward scatter
by a factor of $\sim$ 2) and $M_{\rm BH}-\sigma$ correlations (with a downward scatter
by a factor of $\sim 3-7$, depending on the form adopted for the
dependence of $M_{\rm BH}$ on $\sigma$).}
{Although NGC~3998 cannot be considered as an outlier, 
its location with respect to the $M_{\rm BH}-\sigma$ 
relation conforms with the trend suggesting the presence of a connection
between the {\sl residuals} \rm from the $M_{\rm BH}-\sigma$ correlation
and the galaxy's effective radius. 
In fact, NGC 3998 has one of the smallest values of 
$R_{\rm e}$ among the galaxies with measured $M_{\rm BH}$ (0.85 kpc) 
and it shows a negative residual. 
This suggests that a combination of both $\sigma$ and $R_e$ 
is necessary to drive the
correlations between $M_{\rm BH}$ and other bulge properties, an indication
for the presence of a black holes ``fundamental plane''.}


\titlerunning{The supermassive black hole in NGC 3998}
\maketitle

\section{Introduction}
\label{intro}

It is generally believed that supermassive black holes (SMBHs; $\sim
10^{6}-10^{10}  M_{\odot}$) are a common, if not universal,
feature in the nuclei of nearby galaxies.  Since the discovery of
quasars \citep{schmidt63}, it has been suggested that active galactic
nuclei (AGNs) are powered by mass accretion onto a SMBH
\citep{salpeter64, lynden69}.  This belief, combined with the observed
evolution of the space density of AGNs and the high incidence of
low-luminosity AGN-like activity in the nucleus of nearby galaxies
\citep{heckman80, maoz95, ho97a, ho97b, braatz97, barth98, barth99,
nagar02} implies that a significant fraction of galaxies in the Local
Universe must host black holes (BHs), relics of past activity 
\citep{soltan82, chokshi92, yu02, marconi04, shankar04}.

Supports to these beliefs came from studies of the centers of nearby
early-type galaxies, which revealed that most contain SMBHs 
\citep{Kormendy01, Merritt01b} and that the BH mass
$M_{\rm BH}$ correlates with some properties of the host galaxy, such
as bulge luminosity $L_{\rm bul}$ and mass $M_{\rm bul}$ 
\citep{kormendy95, magorrian98,  
marconi03}, light concentration \citep{graham01} 
and bulge velocity dispersion $\sigma_{\rm bul}$ 
\citep{ferrarese00, gebhardt00a, tremaine02}. The correlation
with the bulge velocity dispersion was thought to be the tightest
having the smallest scatter: rms $\sim$ 0.3 in log $M_{\rm BH}$.
\citet{ferrarese00} argued that, for their selected sample of 12
galaxies, thought to have the most reliable BH mass estimates, the
observed scatter in the $M_{\rm BH}-\sigma_{\rm bul}$ relation
was fully accounted for by the assumed measurement errors, which
implies that there may be no intrinsic scatter in the correlation.
More recently, \citet{marconi03} have shown that when considering
only galaxies with secure BH mass
and bulge parameters determinations all the above correlations have a
similar observed scatter \citep[see also][]{mclure02, erwin04, haring04}. 

The existence of any intrinsic correlations of $M_{\rm BH}$ and host
galaxy bulge properties supports the idea that the growth of SMBHs and
the formation of bulges are closely linked \citep{silk98, haehnelt00}, 
therefore having important implications
for theories of galaxy formation and evolution. Moreover, SMBH mass
estimates inferred via the above correlations, when more direct methods
are unplayable, enter in a variety of important studies spanning from
AGNs physics and fueling to coeval formation and evolution of the host
galaxy and its nuclear black hole.  These correlations need,
however, to be further investigated by increasing the number of
accurate BH mass determinations in nearby galactic nuclei 
to set these correlations on stronger statistical basis.
In particular such a study has the potential to establish the precise
role of the various host galaxy's parameters in
setting the resulting BH mass. 
To date, reliable SMBH detections have been obtained for a limited number of
galaxies ($\sim$ 30, see \citealt{ferrarese05} for a review), with a
bulk of $M_{\rm BH}$ estimates in the range of $10^{7}-10^{9}  M_{\odot}$. 
Furthermore, the coverage of the mass range is
particularly poor at the low-mass end, the sample being strongly
biased against late Hubble types and low luminosity objects. To
add reliable new points to the $M_{\rm BH}-$ host galaxy properties
planes is then a fundamental task for future developments of
astronomical and physical studies, and is the aim of the work
described in this paper.

Spectral information at the highest possible angular resolution is
required to directly measure the mass of SMBHs: 
the radius of the``sphere of influence'' \citep{bahcall76} of 
massive BHs is typically \ltsima 1\arcsec\
even in the nearest galaxies.  Among the so far most widely used
techniques to detect and estimate BHs masses is stellar dynamical
modeling (e.g. \citealt{dressler88, kormendy95, marel98a, gebhardt00a, 
verolme02}), 
but the interpretation of the data is complex 
involving many degrees of freedom and requiring 
data of very high signal-to-noise ratio \citep{valluri04}. 
Radio frequency measurements of H$_{2}$O
masers in disks around BHs, finally, can be applied only to the
small fraction of the disks inclined such that their maser emission is
directed toward us \citep{braatz97}. A more widely
applicable and relatively simple method to detect BHs is based on gas
kinematics (e.g. \citealt{harms94,ferrarese96, macchetto97, barth01}), through
studies of ordinary optical emission lines from circumnuclear gas
disks, provided that the gas velocity field is not dominated by non
gravitational motions. A successful modeling of the gas velocity field 
under the sole influence of the stellar and black hole potential  
is needed to provide {\sl a posteriori} support for a purely
gravitational kinematics. 
The angular resolution of $\sim 0\farcs1$ of
the Space Telescope Imaging Spectrograph (STIS) on-board $\it {HST}$ is
the most suitable to perform such studies. The wealth of unpublished
data contained in STIS archives represents an
extraordinary and still unexplored resource. 
 
Following this chain of reasoning, we performed a systematical search
for unpublished data in the $\it {HST}$ STIS archive with the aim of
finding galaxies candidates to provide a successful SMBH mass measurement. 
 
In this paper we present the results obtained for the active galaxy NGC 3998. 
From the Lyon/Meudon Extragalactic Database (HyperLeda), NGC 3998 is classified
as an early type (S0) galaxy with an heliocentric radial velocity
of 1040 $\pm$ 18 km s$^{-1}$. With ${\it H}_0$ = 75
km s$^{-1}$ Mpc$^{-1}$ and after correction for Local Group infall onto Virgo,
this corresponds to a distance of 17 Mpc and a scale of 83 pc arcsec$^{-1}$.
NGC 3998 is spectroscopically classified as a LINER with broad
H$\alpha$ emission \citep{heckman80, keel83, ho97a, ho97b} 
and no significant broad-line polarization \citep{barth99} therefore 
indicating a probable direct view to a
BLR. Support to this hypothesis comes from studies of
$\it {HST}$ WFPC2 optical images which revealed an unobscured nucleus and the
presence of a bright circumnuclear ionized gas disk \citep{pogge00}.
The nucleus contains a variable compact
flat spectrum radio source \citep{hummel84} displaying a weak
jet-like northern structure \citep{filho02}.  A bright UV source (unresolved
at $\it {HST}$ FOC resolution) is present in the center of
NGC 3998 \citep{fabbiano94}. Its rapidly variable flux \citep{maoz05} 
implies that a significant
fraction of the UV output is contributed by a non-stellar AGN
component. A substantial featureless continuum component is
also observed in the optical \citep{gonzalez04}. The
presence of an AGN in the nucleus of NGC 3998 is further strengthened
by X-ray observations, which revealed the presence of a nuclear source
with a power law spectrum \citep{roberts00, pellegrini00} and 
X-ray and H$\alpha$ luminosity values consistent with
the low luminosities extension of the correlation observed for
Seyferts and QSOs (see \citealt{koratkar95}).  

The paper is organized as follows: in Sec. \ref{obs} we present
HST/STIS data and the reduction that lead to the results described in
Sec. \ref{results}. In Sec. \ref{fitting} we model the observed
H$\alpha$ emission line rotation curve and  we 
show that the dynamics of the circumnuclear gas
can be accurately reproduced by circular motions in a disk when a
point-like dark mass is
added to the stellar potential. 
Our results are discussed in Sec. \ref{discussion},
and summarized in Sec. \ref{summary}.

\section{HST data and reduction}
\label{obs}

\begin{figure*}
\centering
\psfig{figure=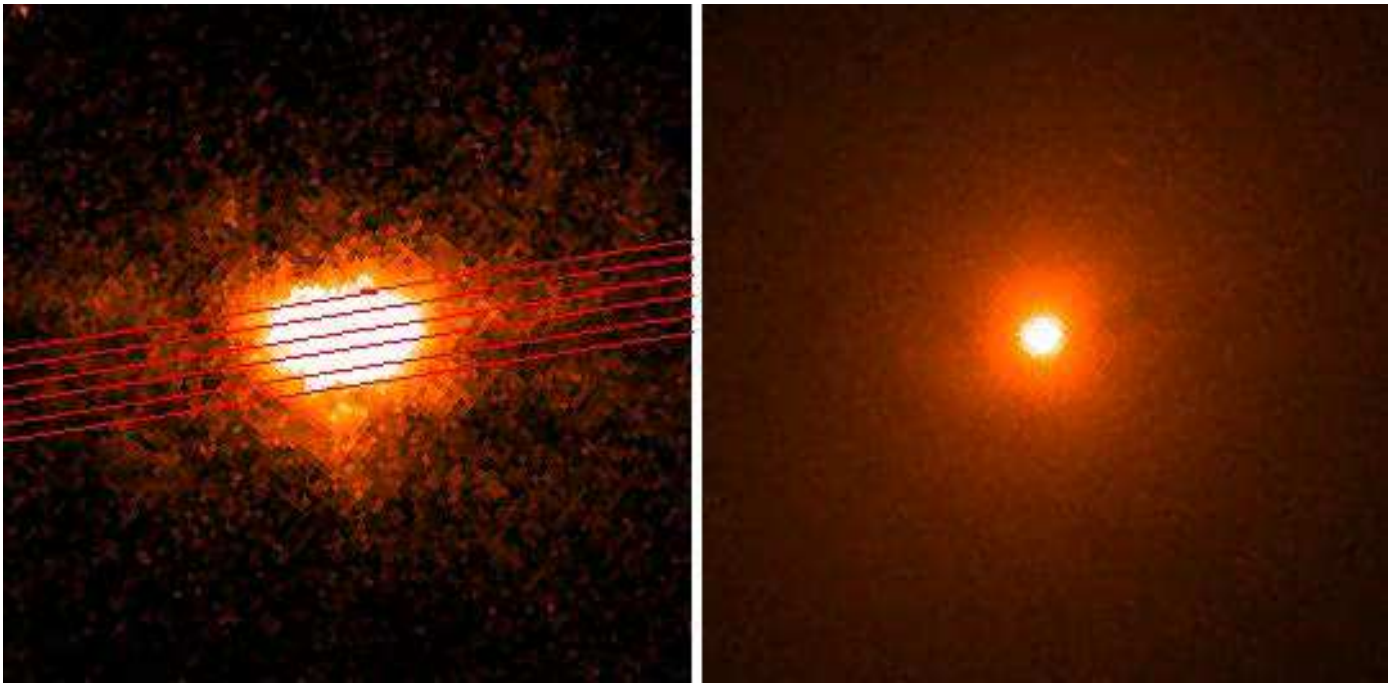,width=1.0\linewidth}
\vskip 0.5cm
\caption{\label{slits} Left) H$\alpha$ image of the central 4\arcsec
$\times$ 4\arcsec of NGC 3998 with superposed the 5 slit
locations. Right) V band (F547M filter) WFPC2/HST image of NGC
3998. North is up, East is left.}
\end{figure*}

NGC 3998 was observed with STIS on {$\it HST$} in 1997 November 01 
with the G750M grating and the 52\arcsec $\times$ 0\farcs1 slit.
Data were acquired at five different slit positions, following a
perpendicular-to-slit pattern with a step of 0\farcs1\ and the central slit
centered on the nucleus. The orientation of the slit was $-81^\circ$ from
North, and the exposure time was 328 s for each position.
The position of the continuum peak had been acquired with the ACQ mode with
two 10 s exposures obtained with the optical long-pass filter MIRVIS.
The five spectra obtained, NUC for the nuclear slit, 
N1-N2 and S1-S2 (from North to South) for
the four off-nuclear, were retrieved from the public archive.
Fig. \ref{slits} shows  the slit locations superposed onto
a narrow band WFPC2, obtained by using the F658N filter, including 
the H$\alpha$+[N II] lines.

The data were obtained 
with a 2x1 on-chip binning of the detector pixels and automatically
processed through the standard {$\it CALSTIS $}
pipeline to perform the steps of bias and dark subtraction, applying
the flat field and combining the two sub-exposures to reject cosmic-ray events.
The data were then wavelength and flux calibrated with conversion to
heliocentric wavelengths and absolute flux units and rectified for the
geometric distortions.
The 2-D spectral image obtained for each slit position has a spatial scale
of 0\farcs0507 pixel$^{-1}$ along the slit, a dispersion of
$\Delta \lambda$ = 1.108 \AA \ pixel$^{-1}$ and a spectral resolution of
$\mathcal{R} \simeq 3000$, 
covering the rest frame wavelength range 6480-7050 \AA.

\begin{figure}
\centering{
\psfig{figure=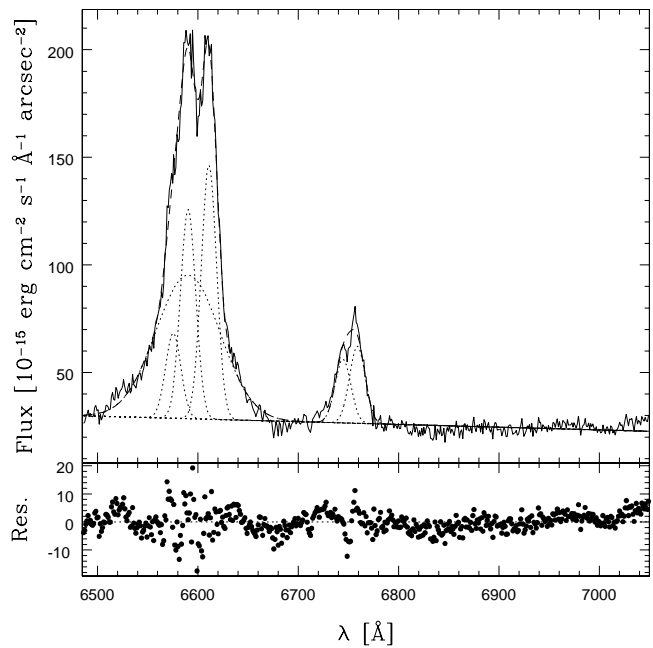,width=1.0\linewidth}}
\caption{\label{profiles} Nuclear spectrum of NGC 3998 showing the presence of
  a broad H$\alpha$ component. The best model spectrum
  (dashed line) is superposed to the data; dotted lines are for the gaussians 
  and linear continuum components of the fit. 
  The residuals from the fit are shown in the lower panel.}
\end{figure}

For each spectrum we selected the regions containing the lines of
interest.  The lines were fitted,  row by row, along the dispersion direction,
together  with a  linear continuum,  with  Gaussian functions  using the  task
SPECFIT in STSDAS/IRAF. All emission  lines present in the spectra (H$\alpha$,
[NII]$\lambda\lambda$6548,6583 and [SII]$\lambda\lambda$6716,6731) were
fitted simultaneously with the same velocity and width 
and with the relative flux of the
[NII] lines kept fixed to 0.334.
Results of the fit are tabulated in Appendix \ref{fitparameters}.

A single Gaussian function for H$\alpha$ line does not produce
an accurate fit to the lines profile in the region $r < 0\farcs2$ 
for three positions of the slit: NUC and the nearest N1 and S1.
Here the lines profile shows a broad base component superposed to the
narrow lines of H$\alpha$ and [N II]
(see Fig. \ref{profiles} for NUC slit), 
indicative of the presence of a Broad Line Region.  
In these regions we performed a fit with two Gaussians components 
for H$\alpha$ line, a narrow and a broad one whose velocity
we kept fixed at all locations. 

In the external regions, where the SNR was insufficient 
the fitting was improved by co-adding two or
more pixels along the slit direction.

\begin{figure}
\centering {
\psfig{figure=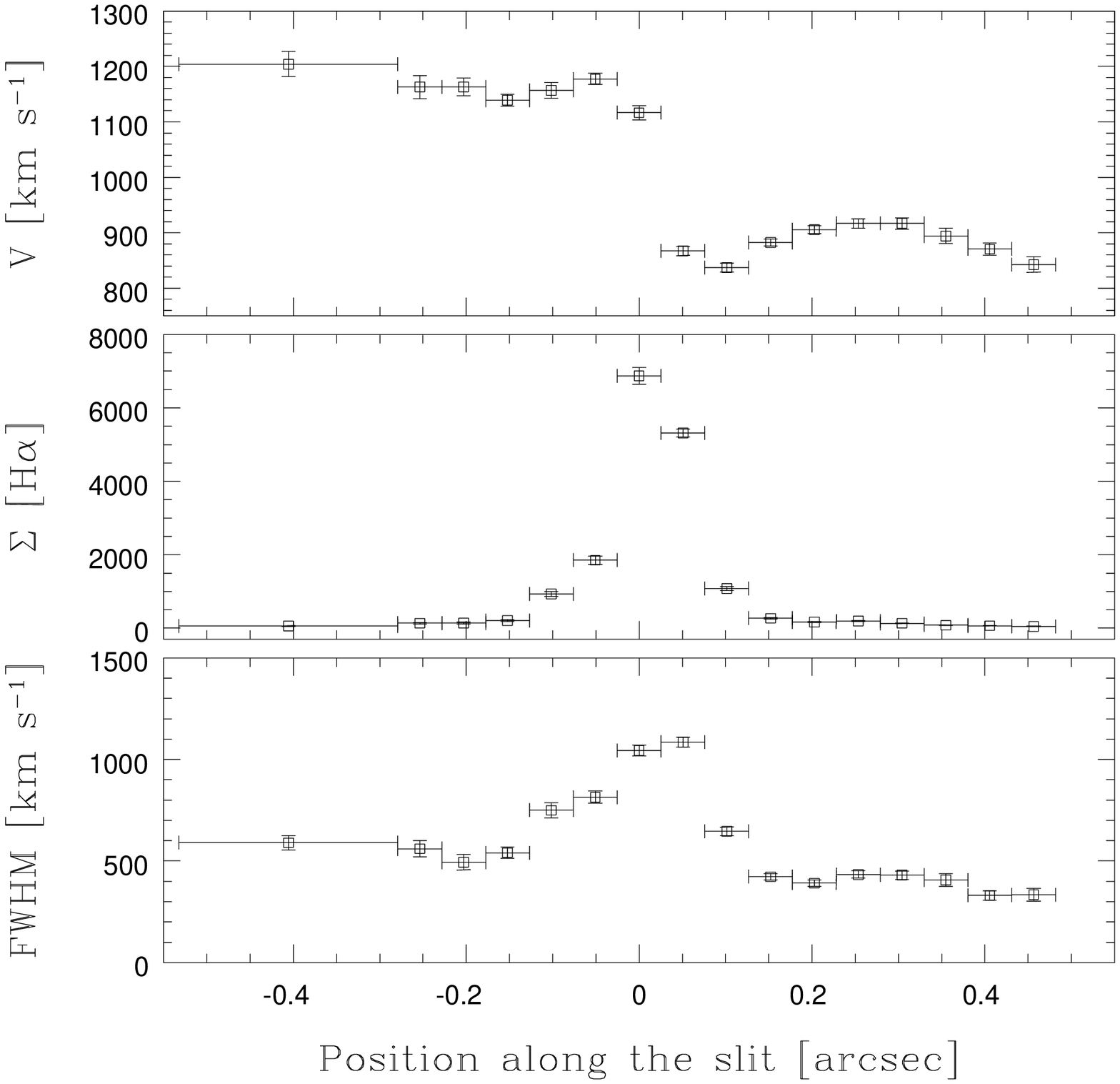,width=1.0\linewidth}}
\caption{\label{nuc} Velocity, surface brightness and FWHM for the narrow
  H$\alpha$ component along the NUC slit.
Surface  brightness   is  in  units  of  10$^{-15}$   erg  s$^{-1}$  cm$^{-2}$
arcsec$^{-2}$.  Positions along  the slit are relative to  the continuum peak,
positive values are NW.  In the  region $r < 0\farcs2$ fits are performed with
a  double  Gaussian  for H$\alpha$  line  to  separate  the broad  and  narrow
components.}
\vskip 0.5cm
\end{figure}

\begin{figure*}
\centerline{
\psfig{figure=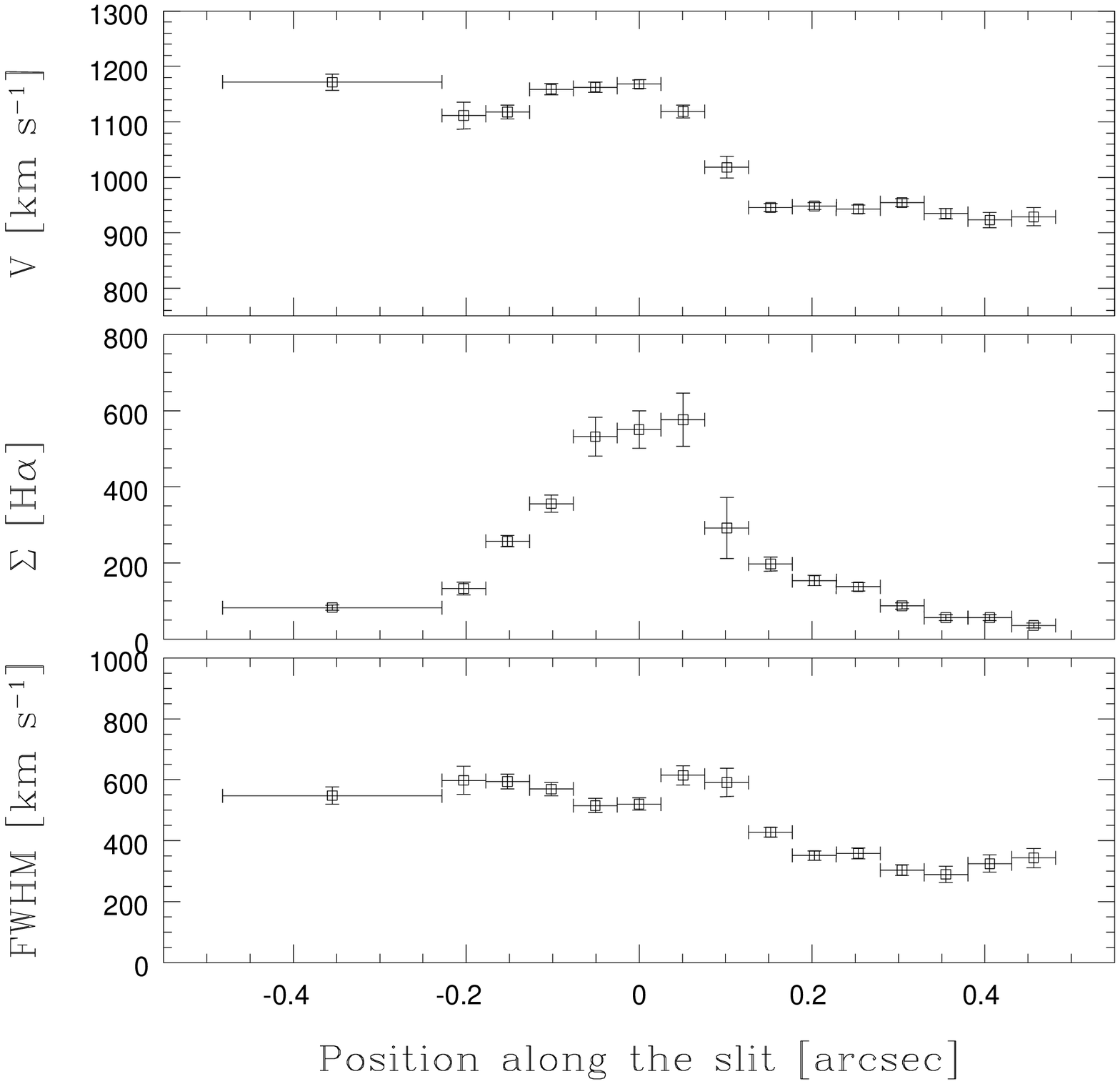,width=0.5\linewidth}
\psfig{figure=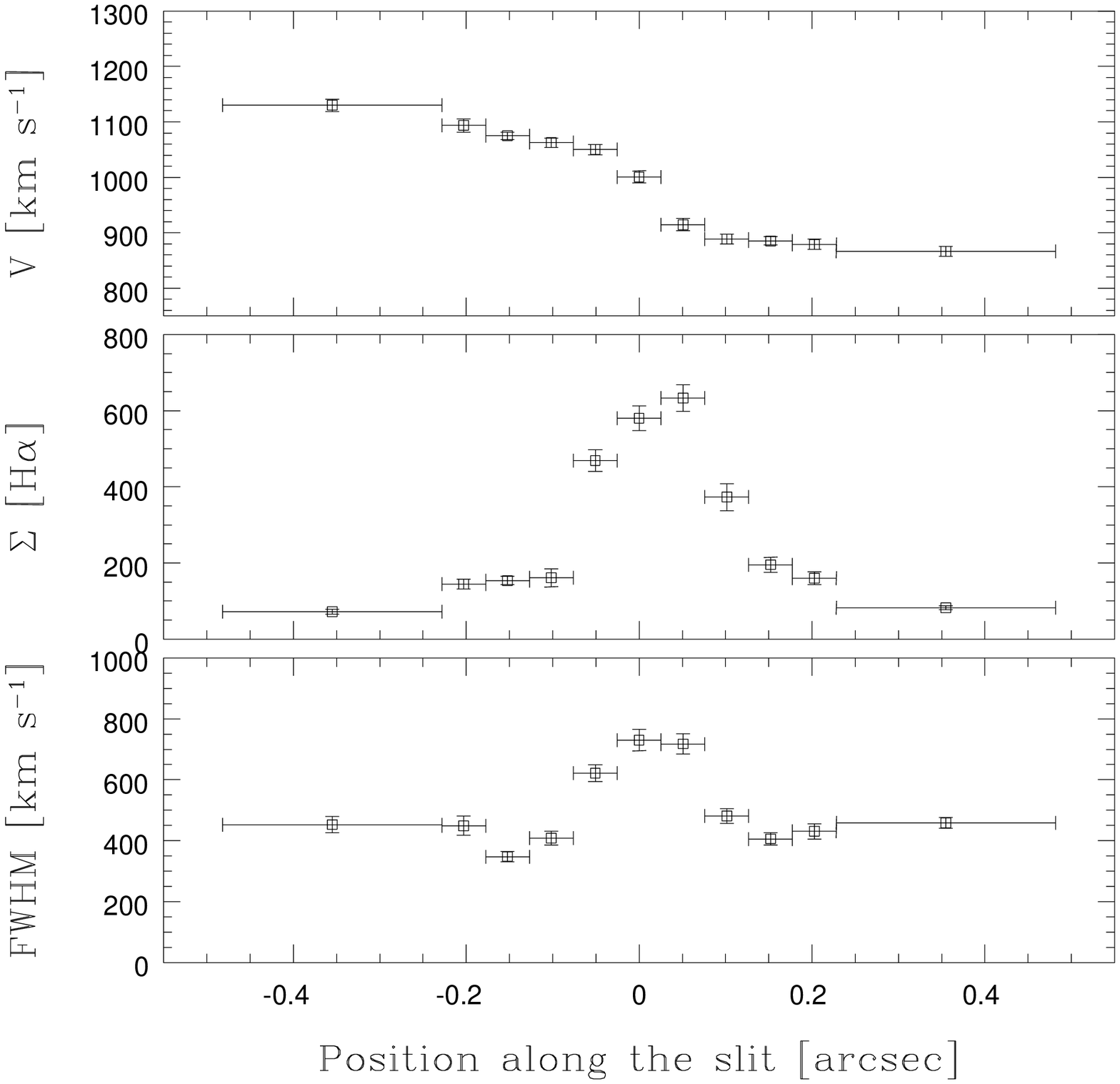,width=0.5\linewidth}}
\caption{\label{near} Same as Fig. \ref{nuc} for the nearest nucleus
slit positions N1 (left panel) and S1 (right).  In the region $r < 0\farcs2$ 
fits are performed with a double Gaussian for H$\alpha$ line
to separate the broad and narrow components.}
\vskip 0.5cm
\end{figure*}

\begin{figure*}
\centerline{
\psfig{figure=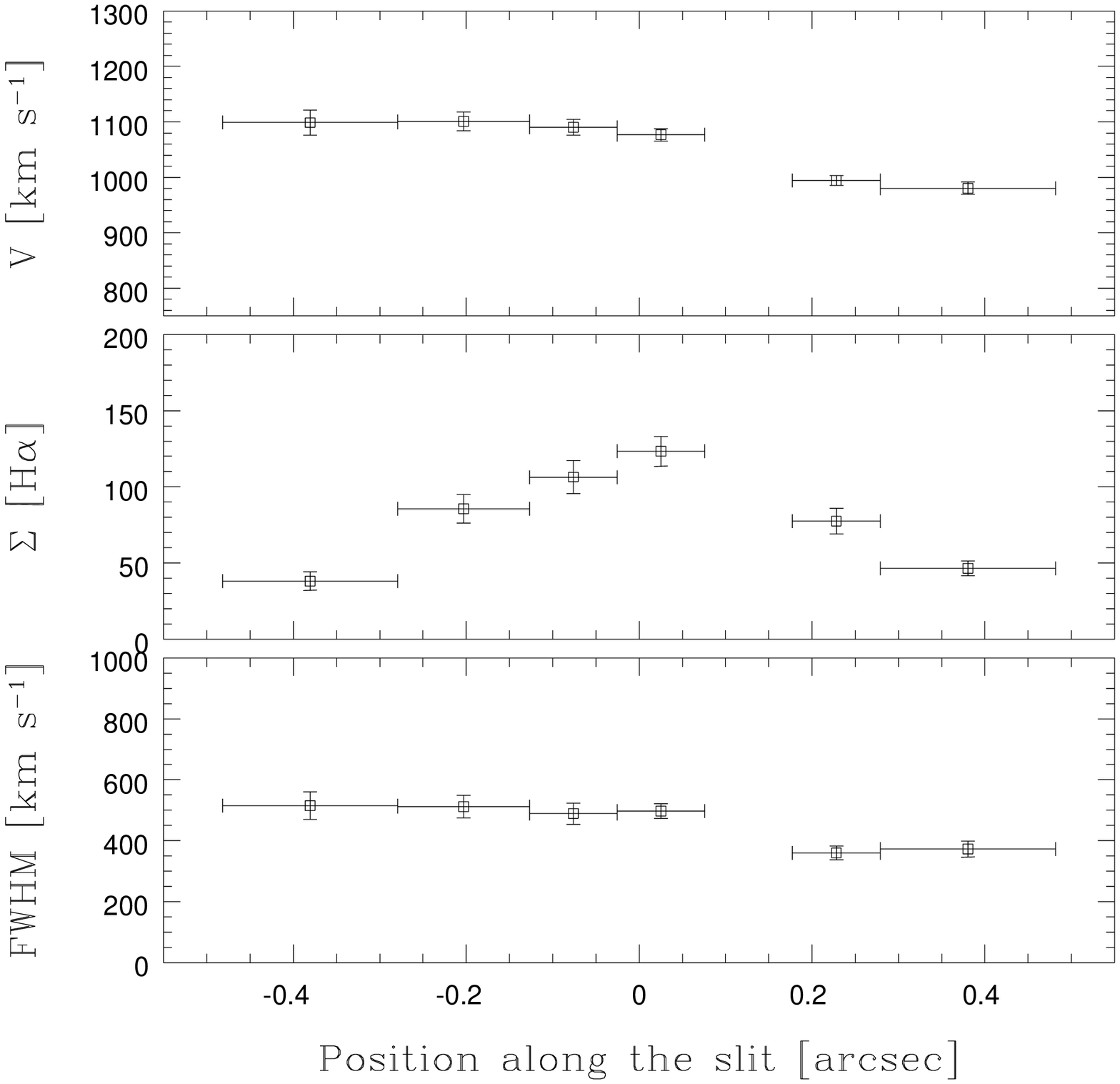,width=0.5\linewidth}
\psfig{figure=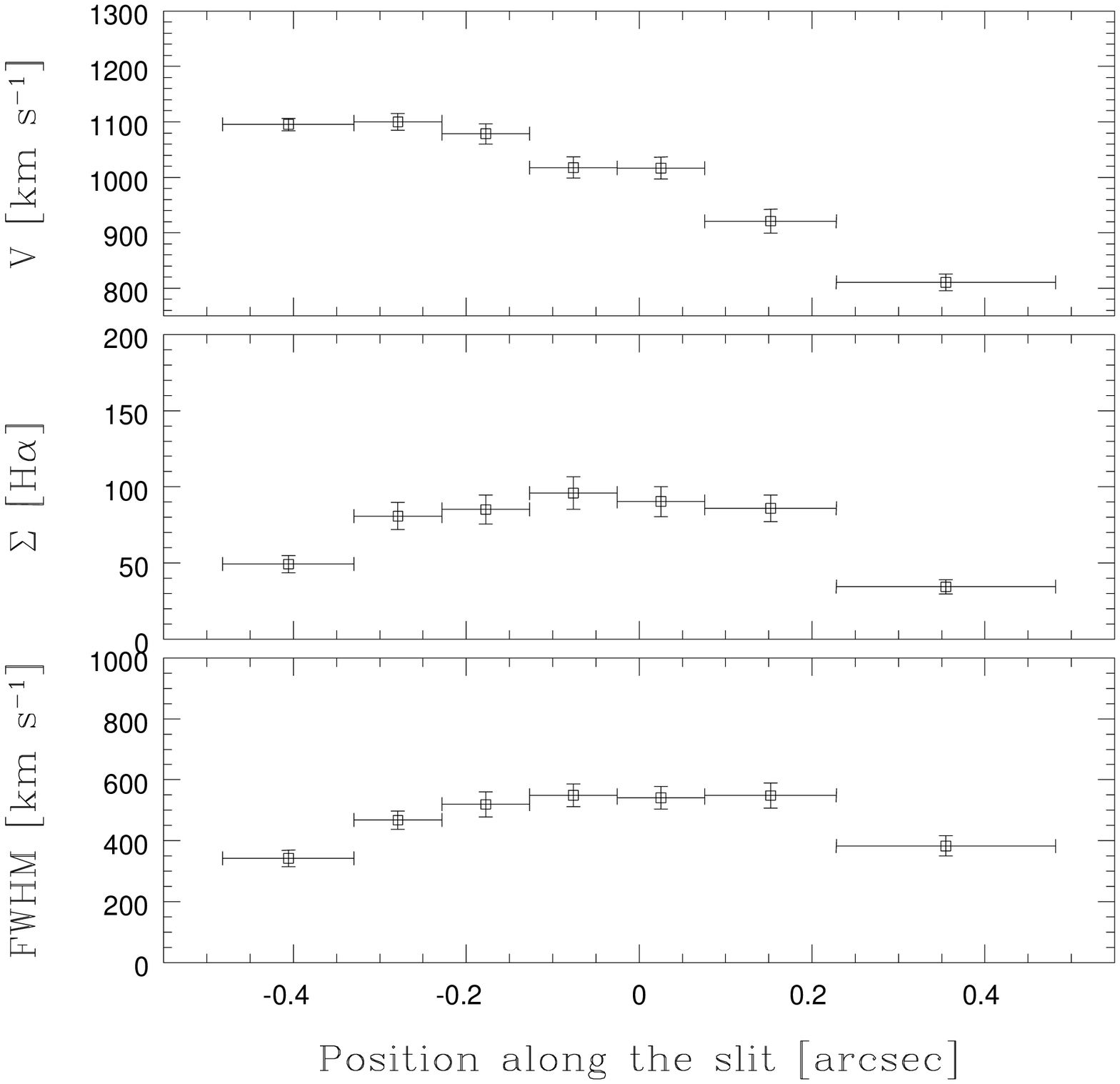,width=0.5\linewidth}}
\caption{\label{far} Same as Fig. \ref{nuc} for the N2 (left panel)
and S2 (right) slit.}
\vskip 0.5cm
\end{figure*}

\section{Results}
\label{results}

The results obtained from the fitting procedure
at the five slit positions are shown in Fig. \ref{nuc} through
Fig. \ref{far} where we show the central velocity, surface brightness
and FWHM for the narrow H$\alpha$ line at each location along the slits. 
Emission is detected
out to a radius of $\sim$ 0\farcs5 corresponding to $\sim$ 40 pc.
The  extension of the line emission  and its
behaviour along the  five slit positions are in agreement with 
the H$\alpha$ image
seen  in  Fig.   \ref{slits},  where  an $\sim$  EST-WEST  elongated  emission
structure is visible. Conversely, the stellar component 
shows an almost circular distribution in the V band continuum.

The  line emission is  strongly peaked  on the  central slit  (Fig. \ref{nuc},
middle panel), rapidly decreasing at larger radii and not showing the 
presence of any emission line knot other than
the central  maximum.  
The velocity curve in the  central slit, NUC, 
has a full amplitude of $\sim 400$ \kms  
and shows a general reflection
symmetry: starting from the center 
the velocity rapidly rises on both sides  by $\sim 200$ \kms
reaching a peak at $r \sim$ 0\farcs1 from the center. 
At larger radii,
the velocity decreases to form a plateau before rising again
at the extremes of the velocity field.
Both the line flux and the line width rapidly decrease from the
nucleus outwards.

The behaviour seen in the off-nuclear slits is qualitatively similar
to that seen at the NUC location, but with substantially smaller
velocity amplitude and, more important, a less extreme
nuclear gradient. Both amplitude and gradient decrease at increasing
distance of the slit center from the nucleus, with a behaviour
characteristic of gas rotating in a circumnuclear disk. 

\section{Modeling the rotation curves}
\label{fitting}

It therefore appears  that an ionized gas  system 
is present in the innermost regions of NGC
3998 with a smooth and regular velocity
field, co-rotating with respect to the  larger scale stellar and
gas disk \citep{fisher97}. 

Our modeling code, described in detail in \citet{marconi03_2},
was used to fit the observed rotation curves. The code computes
the rotation curves of the gas assuming that the gas is rotating
in circular orbits within a thin disk in the galaxy potential.
The gravitational potential has two components: the stellar potential
(determined in the next section), characterized by its mass-to-light ratio
and a dark mass concentration (the black hole),
spatially unresolved at HST+STIS resolution and
characterized by its total mass $M_{\rm BH}$.
In computing the rotation curves we take
into account the finite spatial resolution of HST+STIS,
the line surface brightness distribution and we integrate over
the slit and pixel area. The $\chi^2$ is minimized to determine the
free parameters using the downhill simplex algorithm by \citet{press92}.

\subsection{The stellar mass distribution}
\label{stelle}
In order to assess the contribution of stars to the gravitational potential
in the nuclear region, we
derived the stellar luminosity density 
from the observed surface brightness distribution.

\begin{figure}
\centering
\psfig{figure=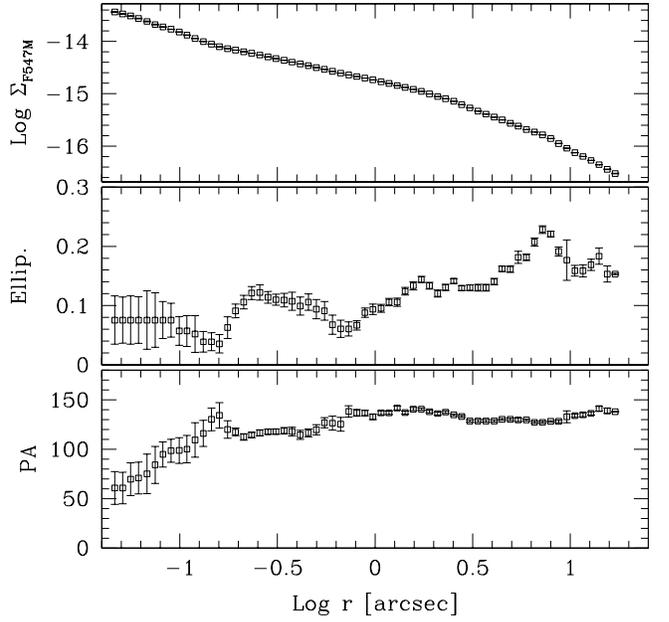,width=1.0\linewidth,angle=0}
\vskip 0.5cm
\caption{\label{ellipse} Results of the isophotes analysis of the V band
image of NGC 3998. Surface brightness is shown in the top panel
(in units of erg s$^{-1}$ cm$^{-2}$ \AA$^{-1}$ arcsec$^{-2}$), the
galaxy's ellipticity and position angle are shown in the middle and
bottom panels respectively.}
\end{figure}

We reconstructed the  galaxy light profile using a WFPC2  F547M (V band) image
retrieved from the public archive (Fig. \ref{slits}). An inspection of
 {$\it HST$} archive images shows that contamination from dust
is negligible, since dust absorption is seen only on relatively 
large scale ($r >$ 2 kpc) while the central regions are apparently 
free from dust (see also \citealt{pogge00}).
We used the IRAF/STSDAS
program   ELLIPSE   to  fit   elliptical   isophotes   to   the  galaxy   (see
Fig. \ref{ellipse}). Excluding the  nuclear regions ($r \leq$ 0\farcs15) that
are dominated by a central compact source, the  ellipticity shows small
variations around a value of $\sim $ 0.15.
The position angle does not show significant variations
being approximately constant at PA = 135$^\circ$.

\begin{figure}
\centering
\psfig{figure=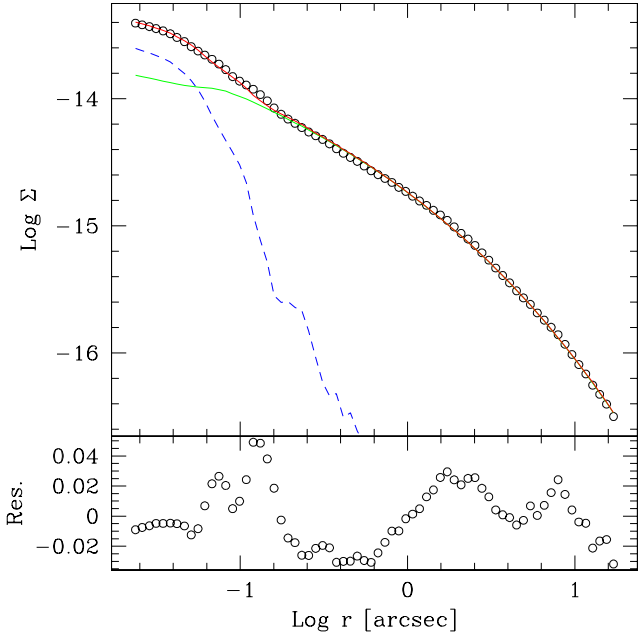,width=1.0\linewidth,angle=0}
\vskip 0.5cm
\caption{\label{prof} Fit to the brightness
profile of NGC 3998 obtained with a S\'ersic law
with an added nuclear point source (dashed line). 
The residuals from the fit are shown in the lower panel.} 
\end{figure}

The nature of the compact nuclear source is crucial 
for the estimate of the stellar mass distribution. In fact, 
if this is associated to light produced by the active nucleus, 
it does not correspond to a stellar mass contribution and should not 
be included in the mass budget. The presence of a bright (and variable)
UV and X-ray source seems to indicate that indeed this is the case. 
We explored in more detail this issue by fitting the brightness profile 
with a S\'ersic law \citep{sersic68} with superposed a point source, 
whose profile was derived from a synthetic Point Spread
Function modeled with TINYTIM. This analysis shows a very good
agreement between the data and the model (see Fig. \ref{prof}).  
This supports the conclusion that the central source is unresolved and
associated to AGN emission. 

The inversion procedure to derive the stars distribution
from the surface brightness
is not unique if the gravitational potential does not have a
spherical symmetry.  
Assuming that the gravitational potential is an
oblate spheroid, the inversion depends on the knowledge of the potential
axial ratio $q$, and the inclination of its principal plane with respect
to the line of sight.
As these two quantities are related by the observed
isophote ellipticity, we are left with the freedom of assuming different
galaxy inclinations to the line of sight.
We performed 
the de-projection adopting for the galaxy inclination the value $i = 40^\circ$,
as given from the HyperLeda database. Due to the small ellipticity of this
galaxy, the precise value of its inclination has only a
marginal effect on the resulting mass distribution.

\begin{figure}
\centering
\psfig{figure=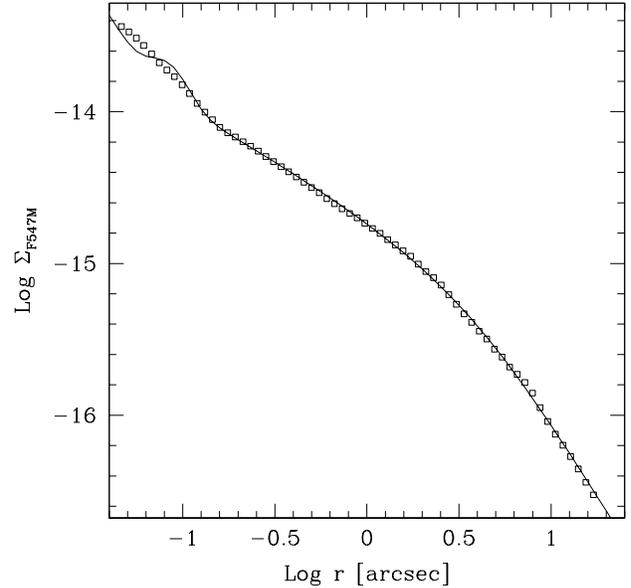,width=1.0\linewidth,angle=0}
\vskip 0.5cm
\caption{\label{mass} Fit to the
surface brightness profile obtained from an oblate spheroid stellar 
density distribution
with an added nuclear point source.}
\end{figure}

Following \citet{marel98}, we assumed an oblate spheroid density
distribution parameterized as:
$$ \rho(m) = \rho_0\left(\frac{m}{r_{\rm b}}\right)^{-\alpha}
\left[1+\left(\frac{m}{r_{\rm b}}\right)^2\right]^{-\beta}
$$ where $m$ is given by $m^2 = x^2+y^2+z^2/q^2$,  $xyz$ is a reference
system with the $xy$ plane corresponding to the principal plane of the
potential and $q$ is the intrinsic axial ratio, and performed the
de-projection adopting for the mass-to-light ratio in the V band the
reference value of 
${\it\Upsilon_V}$ = 1.  A detailed description of the relevant formulas
and of the inversion and fit procedure is presented in
\citet{marconi03_2}.  The best fit obtained is shown in
Fig. \ref{mass} with $\alpha$ = 1.67, $\beta$=0.65 and $r_{\rm b}$
=4.65\arcsec\footnote{Note that with the derived density profile, the stellar mass included
within the HST spatial resolution (0\farcs1) is 
4.0 $\times 10^7  M_{\odot}$,
having adopted a mass-to-light ratio ${\it\Upsilon_V}$ = 6.5  that will be derived in the next Section. Thus it only represents a
fraction of $\sim$ 15 \% of the best fit value for the SMBH mass.}. 
A point source with flux 5.2
10$^{-16}$ erg s$^{-1}$  cm$^{-2}$ \AA$^{-1}$ was added to the extended
luminosity distribution. 

\subsection{Fitting the gas kinematics}

Our modeling code was used to fit the nuclear rotation curves.
The free parameters of the fit are: 
\begin{itemize}
\item
the systemic velocity, $v_{\rm sys}$,
\item
the impact parameter (i.e. the distance between
the nuclear slit center and the center of rotation) $b$,
\item
the position of the galaxy center along the nuclear slit $s_0$,
\item
the angle between the slits and the line of nodes, $\theta$,
\item
the disk inclination $i$,
\item
the mass-to-light ratio of the stellar component, ${\it\Upsilon_V}$,
\item
the black hole mass $M_{\rm BH}$.

\end{itemize}

In a oblate spheroid, the stable orbits of the gas are coplanar with
the principal plane of the potential and it is possible to directly
associate the galaxy inclination and line of nodes with those of the
circumnuclear gas.  However, the potential shape is not sufficiently
well determined by the isophotal fitting down to the innermost regions
of the galaxy and it is possible that a change of principal plane
might occur at the smallest radii, in particular within the sphere
of influence of a
supermassive black hole.  By these considerations we preferred to
leave the disk inclination as a free parameter of the fit.  We then
performed a $\chi^2$ minimization for different values of $i$, namely
$i= 10^\circ,...,80^\circ$, allowing all other parameters to vary
freely.  Due to the sensitivity of the observed line width to the
brightness distribution modeling and to other computational problems
\citep[i.e. a coarse sampling,][]{marconi06} we decided to
initially perform the fit without using the values of the line width.
We will show in Sect. \ref{linewidth} that the inclusion of
the line widths in the modeling code has only a marginal effect on
our results. 
   
To build the synthetic kinematical models the intrinsic line surface 
brightness distribution for the narrow H$\alpha$ line     
had to be obtained for each disk inclination. The observed emission
line surface brightness was modeled for each $i$ value with a
composition of three circularly symmetric Gaussian functions.
The first reproducing the observed central emission peak, the second the
intermediate regions and the third to account for the brightness
behaviour at large radii. The modeling was performed through a
$\chi^2$ minimization, leaving as free parameters the intensity, scale
radius and peak position.
The choice of a particular model for the line  
surface brightness distribution does not affect the final BH mass estimate
provided that the model reproduces the observed line emission within 
the errors. Nevertheless
it has an important effect on the quality of the velocity fit
 \citep{marconi06}. 

\begin{figure}
\centering
\psfig{figure=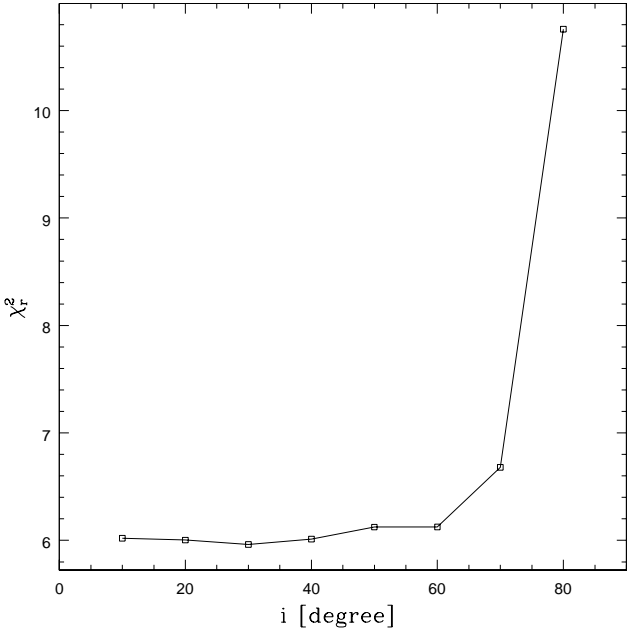,width=1.0\linewidth,angle=0}
\caption{\label{chi} $\chi^2_{\rm r}$ values of the best fit obtained
for each disk inclination.}
\end{figure}

\begin{figure*}
\centerline{
\psfig{figure=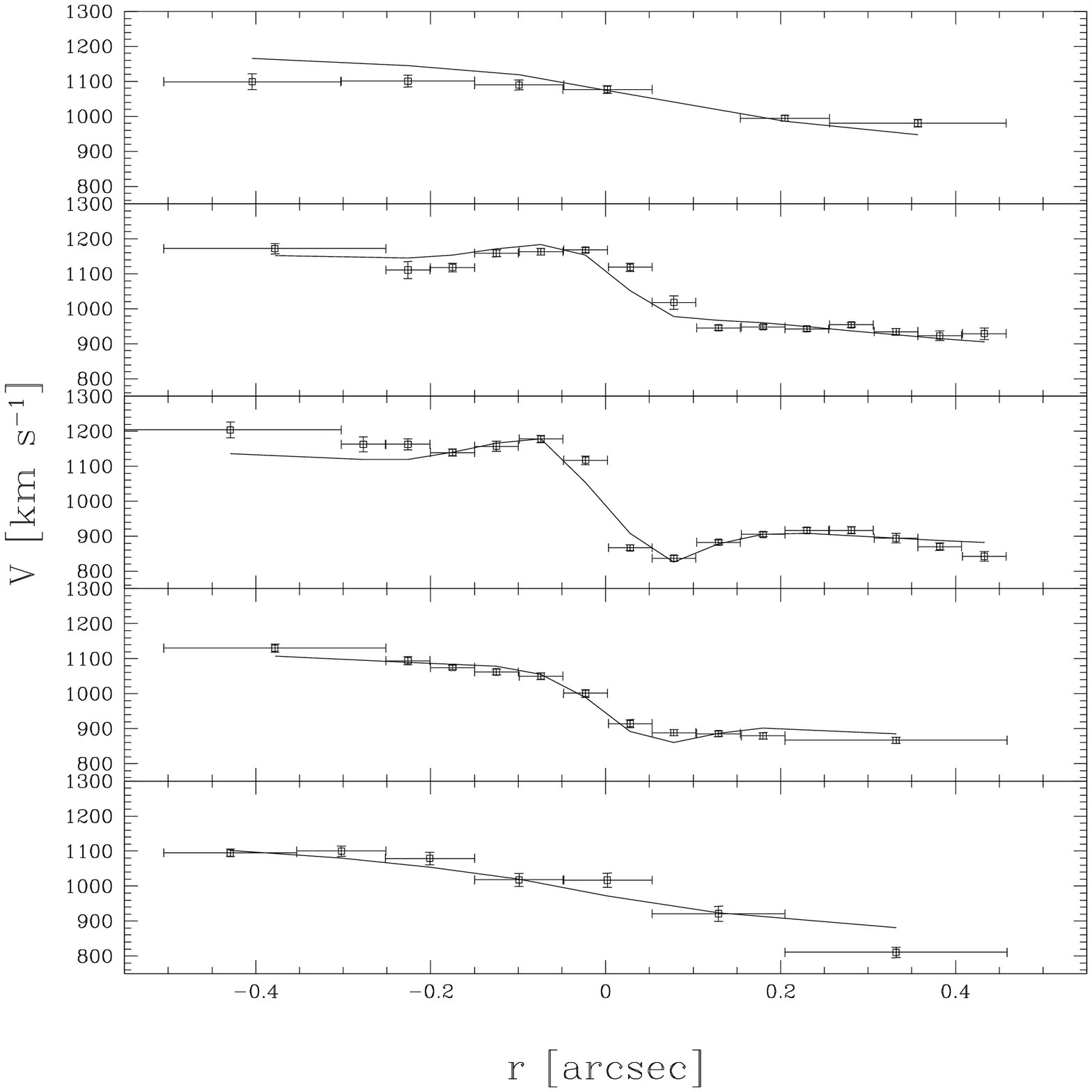,width=0.5\linewidth}
\psfig{figure=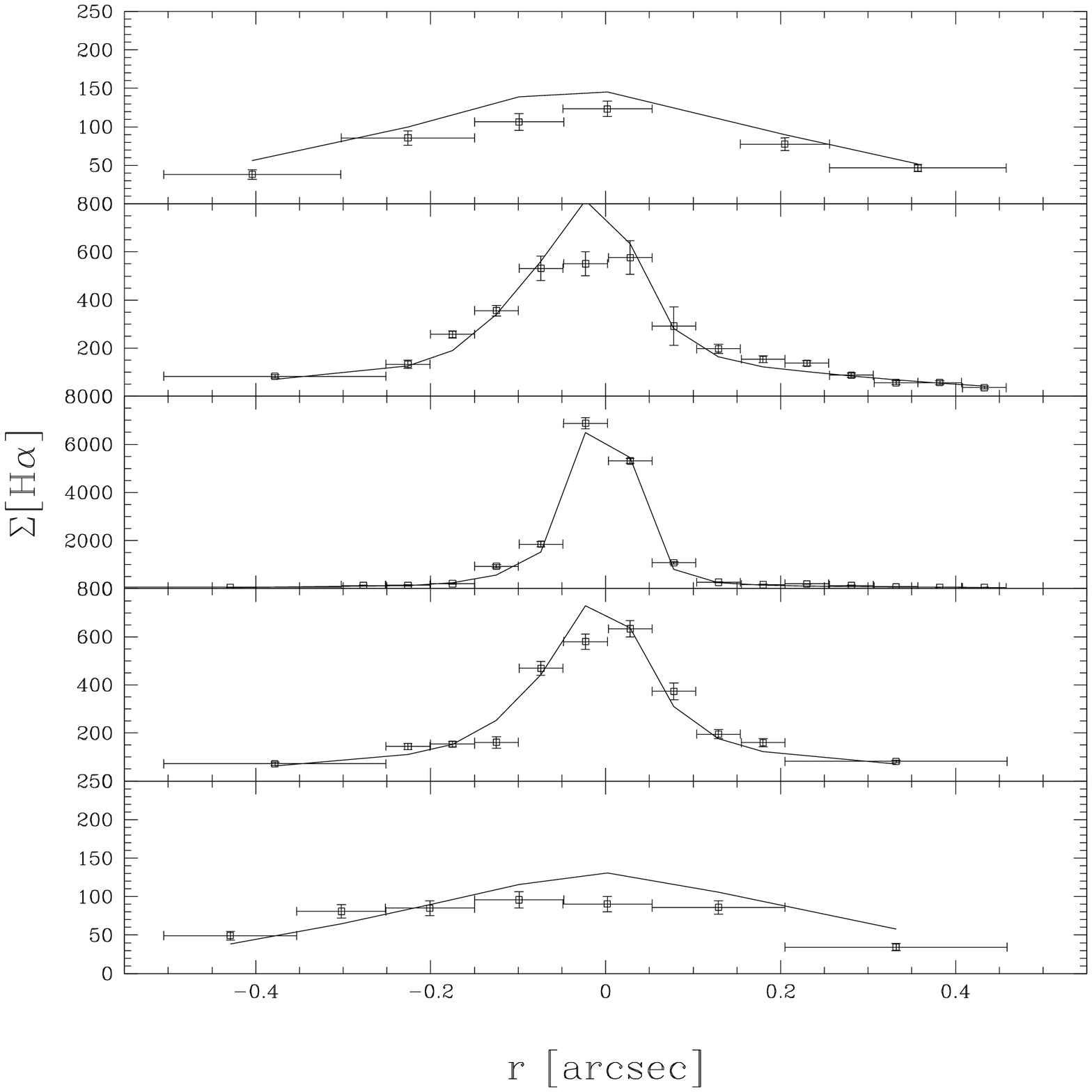,width=0.5\linewidth}}
\caption{\label{fit30} Overall best fit to the rotation curves (left),
assumed line surface brightness distribution (right).  From upper to
bottom panel: N2, N1, NUC, S1 and S2.}
\end{figure*}

The $\chi^2$ reduced values ($\chi^2_{\rm r}$ = $\chi^2$/d.o.f.) of
the best fit to the velocity curves, obtained for each disk
inclination, are shown in Fig. \ref{chi}.  The quality of the fit
depends only very weakly on the assumed disk inclination, with almost
constant values of $\chi^2_{\rm r}$ up to $i= 60^\circ$, while
$\chi^2_{\rm r}$ rapidly increases at larger inclinations.  The
overall best fitting model to our data is obtained for $i= 30^\circ$
for the set of parameters reported in Table \ref{bestfit} and is
presented in Fig. \ref{fit30}.

\begin{figure}
\centerline {
\psfig{figure=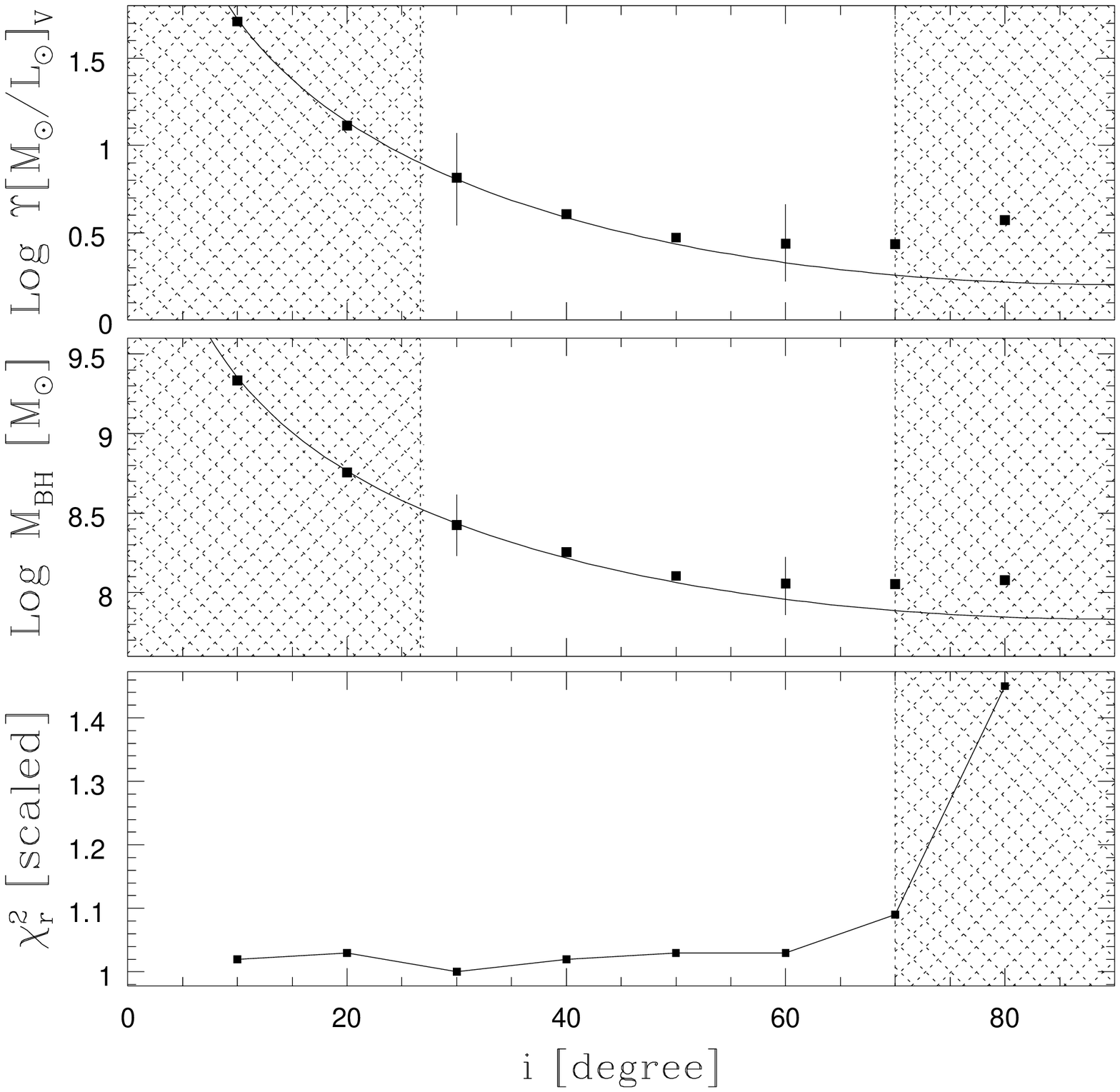,width=1.0\linewidth}}
\caption{\label{gamma} Best fit values of ${\it\Upsilon_V}$ and $M_{\rm BH}$
obtained by varying the disk inclination $i$ (upper and middle
panels). The corresponding $\chi^2_{\rm r}$ are shown in the bottom
panel.  The lines in the upper and middle panels reproduce the
$\propto$ 1/$\sin ^2 i$ dependence of $M_{\rm BH}$ and ${\it\Upsilon_V}$.
Inclinations larger than 70$^\circ$ are excluded at a confidence level
of 2$\sigma$ (shaded region in the right side of the diagrams).
Models with inclinations smaller than 27$^\circ$ are excluded since
they correspond to unacceptably high values of ${\it\Upsilon_V}$ (shaded
region in the left side of the diagrams, bounded by ${\it\Upsilon_V}$ = 7.87
as expected for a Salpeter IMF).}
\end{figure}

The value of minimum $\chi^2_{\rm r}$ is far larger than the value
indicative of a good fit and this is in contrast with the fact that
the curves shown in Fig. \ref{fit30} seem to trace the data points
well.  The reason for this discrepancy is that $\chi^2_{\rm r}$ is not
properly normalized (e.g.  because not all points are independent or
as they do not include the uncertainties in the relative wavelength
calibration for the five slits) and/or imply the presence of small
deviations from pure rotation.  Following \citet{barth01} we then
rescaled the error bars in our velocity measurements by adding in
quadrature a constant error such that the overall best fitting model
provides $\chi^2$/d.o.f. $\sim$ 1.  This is a quite conservative
approach as it has the effect of increasing the final uncertainty on
$M_{\rm BH}$.  The additional velocity error is found to be 28 km
s$^{-1}$.  We rescaled all values of $\chi^2_{\rm r}$ with this
procedure (Fig. \ref{gamma}, bottom panel).  The best fitting models
obtained at varying disk inclination are within the 2$\sigma$
confidence level ($\Delta \chi^2_{\rm r} \leq 0.09$) for $10^\circ
\leq i \leq 70^\circ$.

\begin{table}
\caption{Best fit parameter set}
\begin{tabular}{c c c c c c c c} \hline
 $i$ & $b$ & $s_0$ & $\theta$ & $V_{sys}$ & ${\it\Upsilon_V}$ & $M_{\rm BH}  
 (M_{\odot}$) & $\chi^2_{\rm r}$ \\ \hline 30 & 0.00 & -0.02 & 28 &
1010 & 6.5 & 2.7$\times10^8$ & 5.96 \\ \hline
\end{tabular}
\label{bestfit}
\end{table}

\begin{figure}
\centerline{
\psfig{figure=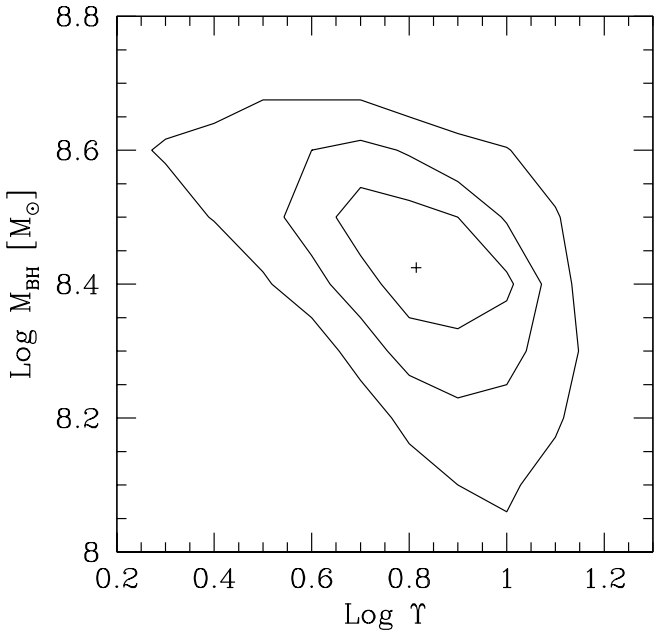,width=1.0\linewidth,angle=0}}
\caption{\label{grid} $\chi^2_{\rm r}$ contours at varying
${\it\Upsilon_V}$ and  M$_{\rm BH}$ at an inclination of $i = 30^\circ$. 
Contours are plotted for confidence levels of 1, 2 and 3 $\sigma$. 
The plus sign marks the overall best fit.} 
\end{figure}

To evaluate the statistical uncertainty associated to the black hole
mass estimate we explored its variation with respect to the parameters
that are more strongly coupled to it, i.e. the mass-to-light ratio
${\it\Upsilon_V}$ and the gas disk inclination $i$.  The uncertainty on
$M_{\rm BH}$ associated to changes in ${\it\Upsilon_V}$ has been estimated
building a $\chi^2_{\rm r}$ grid in the $M_{\rm BH}$ vs. ${\it\Upsilon_V}$
parameter space. At each point of the grid, described by a fixed pair
of $M_{\rm BH}$ and ${\it\Upsilon_V}$ values, we obtained the best fit model
allowing all other parameters to vary freely and derived the
corresponding $\chi^2_{\rm r}$ value (properly rescaled). This enabled
us to build contours of confidence level.  The result of this analysis
at $i = 30^\circ$ is presented in Fig. \ref{grid}.  The 2$\sigma$
ranges of the BH mass and mass-to-light ratio at this disk inclination
are $M_{\rm BH} = 2.7_{-1.0}^{+1.5}\times 10^8  M_{\odot}$ and
${\it\Upsilon_V} = 6.5_{-3.0}^{+5.3}$ respectively. The statistical
uncertainty on $M_{\rm BH}$ is reported in the error bar in
Fig. \ref{gamma}.  We repeated the same analysis for another
representative value of the inclination, $60^\circ$, obtaining similar
fractional uncertainties on the parameters.

The dependence of BH mass on the gas disk inclination closely follows
the expected scaling with a $\propto$ 1/$\sin ^2 i$ law (see
Fig. \ref{gamma}), at least up to $i \leq 60^\circ$.  Since all disk
inclination $i \leq 70^\circ$ yield acceptable fits (within the
2$\sigma$ level), apparently our analysis only provides a lower bound
to the black-hole mass. However, also the mass-to-light ratio
${\it\Upsilon_V}$ increases sharply at low inclinations reaching a value of
50 for $i = 10^\circ$. In fact at smaller inclinations a deeper
potential is needed to maintain the projected velocities seen at the
largest radii where the stellar component dominates.  We then
considered the evolutionary synthesis models for stellar populations
derived by \citet{maraston98} in order to limit ourselves only to
astrophysically acceptable models.  The value of ${\it\Upsilon_V}$
monotonically increases with the galaxy's age reaching, for a Salpeter
initial mass function (IMF), the maximum value in the V band of
${\it\Upsilon_V}$ = 7.87 for an age of 15 Gyr.  
This value corresponds to an inclination of $\sim
27^\circ$ (having used a 1/$\sin ^2 i$ interpolation on the disk
inclination for the dependence of both ${\it\Upsilon_V}$ and $M_{\rm BH}$).  This translates
into an upper bound to the black hole mass (see Fig. \ref{gamma}).
Therefore, although larger black-hole masses are allowed when
considering only the kinematical modeling, they are in reality
unacceptable as they would correspond to a mass-to-light ratio of the
stellar population larger than predicted by the stellar population
synthesis. Furthermore, the adopted upper limit on ${\it\Upsilon_V}$ 
compares favourably with observations. The correlation between
galaxy's luminosity and mass-to-light ratio
\citep[e.g.][]{vandermarel91,cappellari06} predicts a range (in the R band)
of ${\it\Upsilon_R} \sim 3 - 6.5$. This is consistent with an age
of $7 - 15$ Gyr and with the upper limit we adopted in the V band.

By combining the allowed range in inclination with the statistical
uncertainties on $M_{\rm BH}$ associated to variations in ${\it\Upsilon_V}$,
we obtain a global range of acceptable black hole mass of $M_{\rm BH}
= (0.7 - 5.1)\times 10^8  M_{\odot}$ at a 2 $\sigma$ level.

\subsubsection{Line width distribution}
\label{linewidth}

\begin{figure}
\centerline{
\psfig{figure=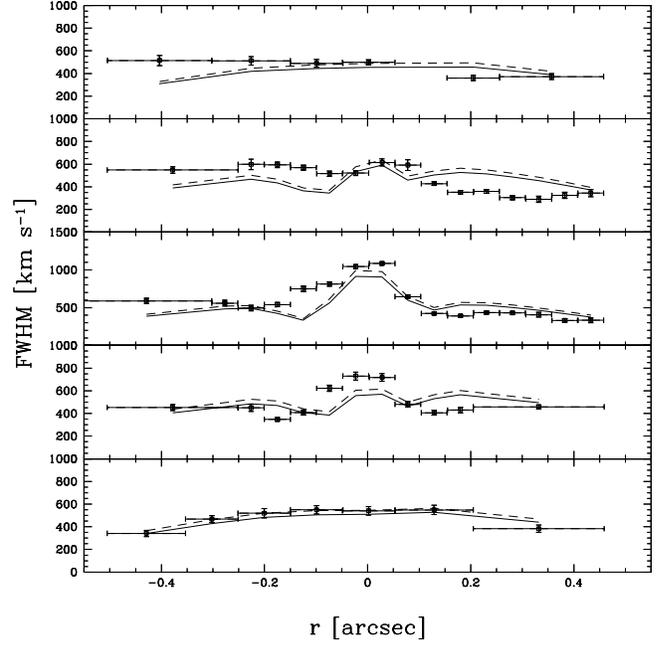,width=1.0\linewidth}}
\caption{\label{fw} Line width distribution expected from the overall
  best fit model (solid line) at $i = 30^\circ$, and derived (dashed
  line) by including the observed line widths in the fitting
  procedure.  From upper to bottom panel: N2, N1, NUC, S1 and S2.}
\end{figure}

At this point of our analysis we tested the influence over the above
results of including the line width in the fitting procedure.
Adopting the best fit parameter set (Table \ref{bestfit}) we obtained the FWHM distribution shown
in Fig.\ref{fw} (solid line).
The observed line widths are acceptably well      
reproduced by the model. Only a slightly underestimate ($\sim$ 20$\%$) of the  
nuclear increase is observable at NUC, while the  
largest deviation from the peak values occurs at the off-nuclear S1.        
Among the line parameters, the line width is the most sensitive to the
brightness distribution modeling and to other computational problems,
i.e. a sub-sampling of the grid used by the numerical code
\citep{marconi06}. Due to this sensitivity only weak constraints on
the BH mass estimate can be derived by its inclusion in the fitting
procedure.  Therefore, a slightly underestimate of the peak observed
values does not invalidate the derived fitting model.

Despite the goodness of the result, we repeated the $\chi^2$
minimization at $i = 30^\circ$, this time by including the observed
line widths in the modeling procedure.  The best fit obtained is shown
in Fig. \ref{fw} (dashed line). Only a small improvement of the match
with observed values is obtained, while the model velocity curves
remain substantially unchanged respect to the previous result.
Furthermore, BH mass and mass-to-light ratio values do not change
significatively, resulting $M_{\rm BH} = 3.1\times 10^8  M_{\odot}$
and ${\it\Upsilon_V}$ = 5.6.
The modeling code assumption that the nuclear gas is in a thin,
circularly rotating disk is then verified through the satisfactory
good match of observed and model line width distributions. Indeed, the
nuclear rise of the line width is well accounted for as unresolved
rotation by the fitting model.

\section{Discussion}
\label{discussion}

Our model fitting of the nuclear rotation curves of NGC 3998 indicates
that the kinematics of gas in its innermost regions can be
successfully accounted for by circular motions in a thin disk when a
point-like dark mass (presumably a supermassive black hole)
of $M_{\rm BH} = 2.7_{-2.0}^{+2.4}\times 10^8  M_{\odot}$ is added to the 
galaxy potential. 

Let us explore how this mass determination is connected with the
properties of the host galaxy.  We compare our BH mass estimate with
the known correlations with host spheroid (bulge) mass
\citep{marconi03} and with the stellar velocity dispersion
\citep{tremaine02,ferrarese05}.

Following \citet{marconi03}, we used the virial mass ($M_{\rm vir}$ =
3$R_{\rm e}\sigma_{\rm bul}^{2}/G$) to determine the bulge mass of NGC
3998. 

The effective radius of the bulge, $R_{\rm e}$, has been
estimated as the weighted average of the determinations by
\citet{fisher96} (11\arcsec) and \citet{sanchez04} (9\farcs6), 
from which we derived $R_{\rm e}$ =
10\farcs2 $\pm$ 0\farcs6 (0.85 $\pm$ 0.05 kpc).  

We found three optical determinations of
the stellar velocity dispersion for NGC 3998 in the literature. The
values are 297 \kms \citep[ 2\arcsec $\times$ 4\arcsec\
slit]{fisher97}, 314 $\pm$ 20 \kms \citep[ 3\arcsec $\times$ 12\arcsec\
slit]{tonry81}, 333 $\pm$ 22
\kms \citep[ 1\farcs5 $\times$ 2\farcs2 slit]{nelson95} respectively
\footnote{\citet{bertola84} measured $\sigma$ = 350 $\pm$
40 \kms, but, because of saturation,
their data gave no results on stellar kinematics in the
innermost central regions ($<$ 2\arcsec).}. 
We adopted the value from HyperLeda database
$\sigma_{\rm star}$ = (305 $\pm$ 10) km s$^{-1}$.
The velocity dispersions in the catalogue are
mean values standardized to a circular aperture of radius $r_{\rm ap}$ = 0.595
h$^{-1}$ kpc \citep[see][]{golev98}.  At the distance of NGC 3998 this
radius corresponds to an aperture of 9\farcs6 of radius, very close
to the value of the bulge effective radius for NGC 3998. 

\begin{figure}
\centering
\psfig{figure=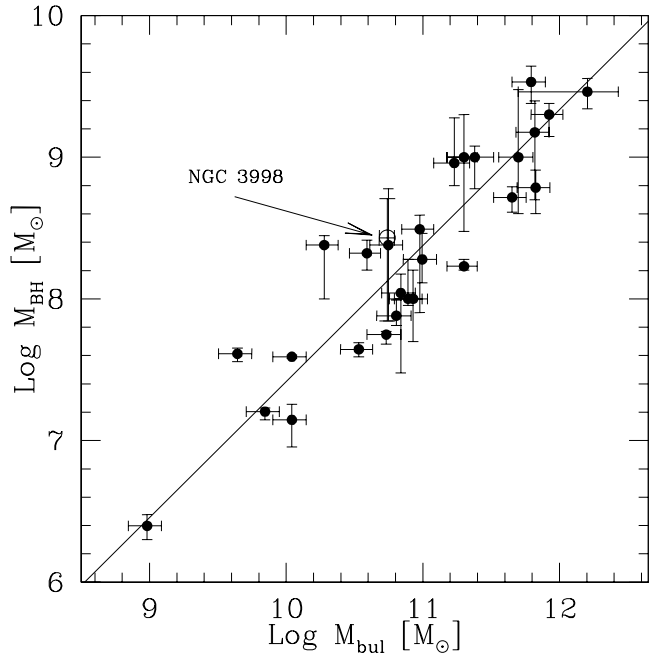,width=1.0\linewidth,angle=0}
\caption{\label{Marconi} $M_{\rm BH}$ vs bulge mass from
  \citet{marconi03} with the best fit obtained from a bisector linear
  regression analysis (solid line). The position of NGC 3998 (empty circle) is
  indicated by an arrow.}
\end{figure}

Assuming that this
value of $\sigma_{\rm star}$ is a good approximation for $\sigma_{\rm
bul}$, we obtained for the bulge mass the value $M_{\rm bul} = (5.5
\pm 0.7) \times 10^{10}  M_{\odot}$. Using this estimate of $M_{\rm
bul}$ and the correlation of \citet{marconi03} which considers only
``secure'' BH mass determinations (i.e. BH with resolved sphere
of influence: 2$R_{\rm sph}/R_{\rm res}>1$, with $R_{\rm res}$ the
spatial resolution of the observations) the expected $M_{\rm BH}$ for
NGC 3998 is 1.3 $\times 10^{8}  M_{\odot}$, in excellent agreement,
within a factor of 2, with our determination (see Fig.\ref{Marconi}).
We also note that the value
of the BH sphere of influence radius in NGC 3998 ($R_{\rm sph} = GM_{\rm
BH}/\sigma_{\rm star}^{2}$) is $R_{\rm sph} \sim$ 13 pc
(0$\farcs16$). This implies  a well resolved BH sphere of influence at the HST
resolution ($\sim 0\farcs1$) for our
$M_{\rm BH}$ determination (2$R_{\rm sph}/R_{\rm res} \simeq$ 3.2).

\begin{figure}
\centering
\psfig{figure=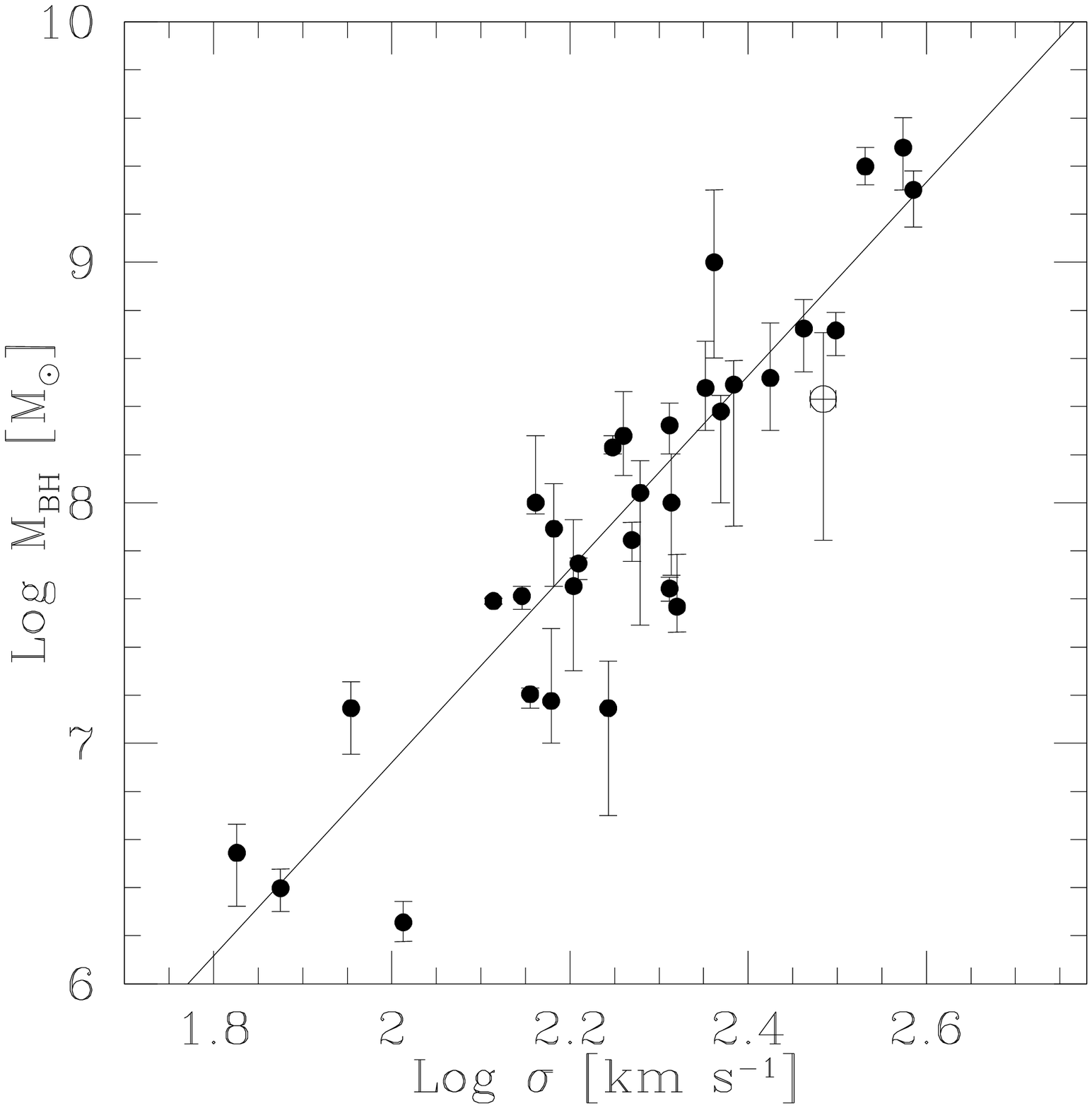,width=0.9\linewidth,angle=0}
\centering
\psfig{figure=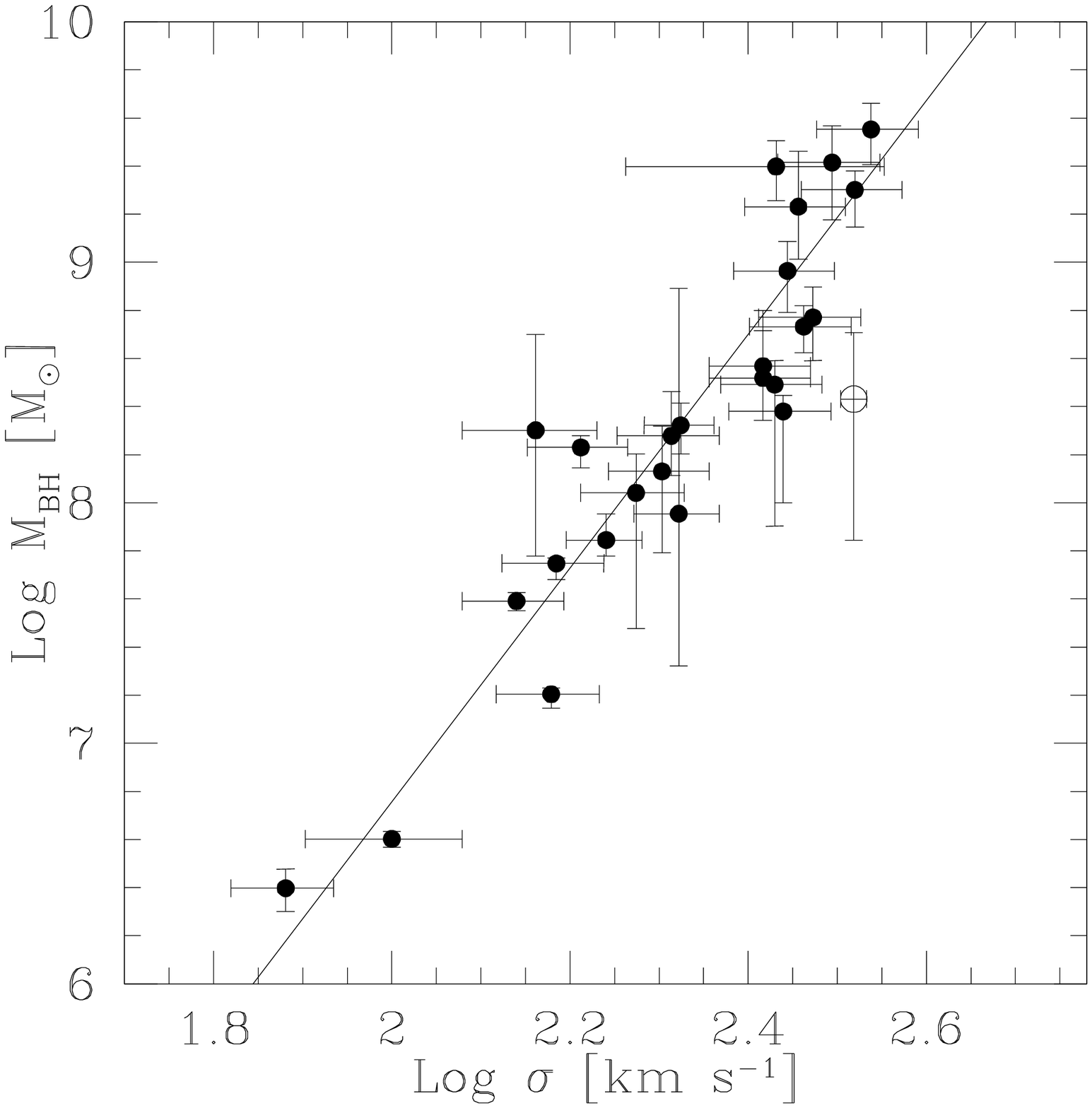,width=0.9\linewidth,angle=0}
\caption{\label{Ferrarese}$M_{\rm BH}$ vs central stellar velocity dispersion from \citet{tremaine02} 
(upper panel) and from
\citet{ferrarese05} (lower panel). Solid lines indicate the best
fits for the two correlations.
The position of NGC 3998 is shown by an empty circle.}
\end{figure}

Concerning the correlations between BH mass and central stellar
velocity dispersion, adopting the correlation parameters
estimated by \citet{tremaine02} the expected BH mass for NGC 3998 is
7.4 $\times 10^8  M_{\odot}$ 
(see Fig.\ref{Ferrarese}), a
factor $\sim$ 2.7 higher than our estimate. However, NGC 3998 
cannot be considered as an outlier from the $M_{\rm BH}$ vs.
$\sigma$ correlation given the 
errors in the estimates of $M_{\rm BH}$, 
$\sigma_{\rm star}$, and of the best fit parameters describing   
$M_{\rm BH}-\sigma$ relation, and its
intrinsic scatter ($0.25-0.3$ in log $M_{\rm BH}$).
We can also compare our measurement 
with the expectations of the $M_{\rm BH}-\sigma$ relation 
recently derived by \citet{ferrarese05}. 
We then normalized the central velocity dispersion to
an aperture of radius equal to 1/8 of $R_{\rm e}$, following the
method introduced by \citet{jorgensen95}, and derived $\sigma_{\rm R_{\rm
e}/8} = 330 \pm 11$ \kms. 
This form of the correlation predicts a BH mass of
1.9 $\times 10^{9}  M_{\odot}$, a factor 
of 7 higher than our estimate, a larger discrepancy 
than the one found adopting the Tremaine et al. values.  

\begin{figure}
\centering
\psfig{figure=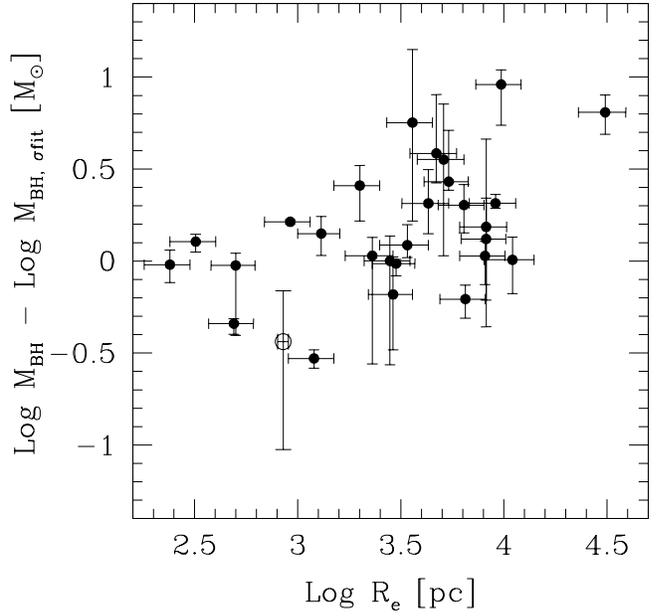,width=1.0\linewidth,angle=0}
\caption{\label{residuals} Residuals from the $M_{\rm BH}$ vs $\sigma$
correlation (in the Tremaine's et al. form) 
reported against the galaxy's effective radius $R_{\rm e}$. 
The position of NGC 3998 is indicated by an empty circle.}
\end{figure}

\citet{marconi03} showed with a 
partial correlation analysis that $M_{\rm BH}$ is separately
significantly correlated both with $\sigma$ and $R_{\rm e}$. 
This is clearly shown by the
residuals of the $M_{\rm BH}-\sigma$ correlation against $R_{\rm e}$ that show
a weak, but significant, correlation (reproduced here in Fig. \ref{residuals}).
The new measurement of the black hole mass in NGC 3998 supports
this idea. In fact, NGC 3998 has one of the smallest values of 
$R_{\rm e}$ among galaxies with measured $M_{\rm BH}$ (0.85 kpc) 
and it shows a negative residual from the $M_{\rm BH}-\sigma$ correlation. 
Recently \citet{Capetti05} found a similar result, but in the opposed
sense, considering 
the Seyfert galaxy NGC 5252: a large effective radius (9.7 kpc)
corresponds in this galaxy to a large positive residual.
This confirms that a combination of both $\sigma$ and $R_{\rm e}$ 
is necessary to drive the
correlations between $M_{\rm BH}$ and other bulge properties, an indication
for the presence of a black holes ``fundamental plane''.

\section{Summary and conclusions}
\label{summary}

We have presented results from a gas kinematics study in the nucleus
of the nearby S0 active galaxy NGC 3998. The analyzed data were
retrieved from archival HST/STIS long-slit spectra. We performed an
analysis of the H$\alpha$, [N II]$\lambda\lambda$6548,6583 and
[SII]$\lambda\lambda$6716,6731 emission lines profiles to derive the
map of the gas velocity field.  The nuclear velocity curves show a
general reflection symmetry and are consistent with the presence of
gas in regular rotation. We used our modeling code to fit the observed
H$\alpha$ surface brightness distribution and velocity curve. The
dynamics of the rotating gas can be accurately reproduced by motions
in a thin disk when a compact dark mass of $M_{\rm BH} =
2.7_{-2.0}^{+2.4}\times 10^8  M_{\odot}$, very likely a supermassive
black hole, is added to the stellar mass component. This result is
also supported by the satisfactory good match of observed and model
line width distributions.  Furthermore, the black hole in NGC 3998 has
a sphere of influence radius of $\sim$ 13 pc (0$\farcs16$). At the
high HST spatial resolution, this value of $R_{\rm sph}$ implies a
resolved BH sphere of influence for our $M_{\rm BH}$ determination
(2$R_{\rm sph}/R_{\rm res} \simeq$ 3.2).

For what concerns the connections of this BH mass estimate with the
properties of the host galaxy, the $M_{\rm BH}$ value for NGC 3998 is
in excellent agreement (within a factor of 2) 
with the $M_{\rm BH}-M_{\rm bul}$ correlation
between BH and host bulge mass. The black hole mass
predicted by the $M_{\rm BH}-\sigma_{\rm bul}$ correlation 
is a factor of 2.7 larger than our measure, adopting the relation
found by \citet{tremaine02}, or a factor of 7 using the most recent 
parameterization by \citet{ferrarese05}. However, NGC 3998 
cannot be considered as an outlier from the $M_{\rm BH}$ vs.
$\sigma$ correlation considering the 
errors in the estimates of $M_{\rm BH}$ and  
$\sigma_{\rm star}$, and both the uncertainties in the 
determination of the $M_{\rm BH}-\sigma$ relation as well as its 
scatter.  

Nonetheless, the lower-than-expected value for 
the mass of the black hole hosted by NGC 3998 strengthens the 
presence of a connection between the residuals 
from the $M_{\rm BH}-\sigma$ relation and the galaxy's effective
radius. In fact, NGC 3998 has one of the smallest values of 
$R_{\rm e}$ among galaxies with measured $M_{\rm BH}$ 
and it shows a negative residual. We also recently showed that the
opposite is true for the Seyfert galaxy NGC 5252: a large effective radius 
corresponds in this galaxy to a large positive residual.

Apparently only with a combination of both $\sigma$ and $R_{\rm e}$ 
it is possible to account for the 
correlations between $M_{\rm BH}$ and other bulge properties, an indication
for the presence of a black holes ``fundamental plane''.
Clearly, only by further increasing the number of direct black hole 
measurements
it will be possible to base these conclusions on a stronger
statistical foundation.

\begin{acknowledgements}
We would like to thank the referee, Laura Ferrarese, for her useful
comments and suggestions.
\end{acknowledgements}

\begin{appendix}
\section{Fitting parameters for STIS spectra.}
\label{fitparameters}
\begin{table*}
\caption{Fitting parameters for the narrow emission line spectra at each
  location (x coord.) along the slit. The relative flux of the [NII] lines was
  held fixed to 0.334, so only the [NII]$\lambda$6583 is listed. From upper to 
  bottom table: NUC, N1, S1, N2 and S2 positions.}
\begin{tabular}{c c c c c c c c c c c c c} \hline
$x$   &    $V$    &   $dV$  &  $FWHM$   & $dFWHM$ &  $\Sigma(H\alpha)$ &
  $d\Sigma(H\alpha)$ & $\Sigma(NII)$  & $d\Sigma(NII)$
  & $\Sigma(SII)$  & $d\Sigma(SII)$ & $\Sigma(SII)$ &
  $d\Sigma(SII)$ \\
  &    &     &           &         &           &         & ($\lambda$6583) & ($\lambda$6583)  & ($\lambda$6716) &
  ($\lambda$6716) & ($\lambda$6731) & ($\lambda$6731)    \\ \hline 
 591-595  &   1204   &    23   &   590   &
   35 &    61   &    7 &  77  & 6 & 25 & 5 &
 24 & 6 \\
   596    &   1163   &    21      &   561   &
 39  &    135   &    16  & 161 & 13 & 42 &
 10 & 63 & 11  \\
   597    &   1163   &    16      &   495   &
 38  &    143     &    24  & 198 & 24 & 89 &
 9 & 77 & 10 \\
   598    &   1139   &    11      &   541   &
 29 &    208   &    26  & 288 & 25 & 142 &
 10  & 150 & 11  \\
   599    &   1157   &    14      &   750   &
 37  &    934   &    71  & 1060 & 72 & 262 &
 25 & 347 &  30 \\
   600    &   1178   &    10       &   814   &
 30 &    1853    &    109 & 2257 & 113 & 577 &
 32 & 701 & 40  \\
   601    &   1117   &    13       &   1044    &
 28 &    6873    &    230   & 6210 & 168 & 883
 & 78 & 1647 & 93 \\
   602    &   867  &     8       &   1085    &
 23 &    5315    &    107  & 4187 & 94 & 766 &
 31 & 1096 & 37  \\
   603    &   837  &     8       &   646   &
 21 &    1081    &    55 & 1140 & 51 & 311 &
 16 & 387 & 20 \\
   604    &   882  &     6      &   422   &
 16 &    269   &    25 &  325 & 20 & 158 &
 8 & 156 & 10    \\
   605    &   905  &     7      &   391   &
 16 &    169     &    18 & 224 & 15 & 117 &
 7 & 129 & 9 \\
   606    &   917  &     8       &   434     &
 18  &    200   &    14  & 220 & 12 & 107 &
 8 & 110 & 9 \\
   607    &   917  &    10       &   431   &
 21 &    134   &    11 & 150 & 10 & 79 &
 6 & 80 & 8 \\
   608    &   894  &    14       &   407   &
 32 &    82   &    9 & 93 & 6 & 44 & 7
 & 41 & 8 \\
   609    &   870  &    11       &   331   &
 23 &    67   &    8 & 78 & 8 & 39 &
 5 & 37 & 6   \\
   610    &   842  &    14    &   333   & 
32 &    51   &    8 & 54 & 7 & 39 &
 5 & 16 & 6 \\ \hline
 592-596  &  1172    &   15        &  548    &
   29  &   83    &   6  & 387 & 5 & 167 & 4 &
 152 & 5 \\
   597    &  1111    &   24        &  598    &
 46  &   133    &   17 & 161  & 13 & 43 & 11 &
 59 & 12  \\
   598    &  1118    &   12         &  594    &
 24 &   257    &   15 & 268 & 12 & 65 & 9 &
 92 & 10  \\
   599    &  1159    &   10         &  569    &
 22 &     356    &   22 & 421 & 22 & 131 & 9 &
 184 & 11    \\
   600    &  1163    &   9         &  516    &
 24 &   532    &   51 & 553 & 41 & 172 & 15 & 216
 & 18 \\
   601    &  1168    &   8         &  521    &
 21 &   551    &   49 & 622 & 40 & 215 & 13 &
 251 & 15      \\
   602    &  1119    &   12         &  615    &
 32 &   576    &   70 & 585 & 53 & 214 & 12 &
 211 & 14    \\
   603    &  1018    &   20         &  592    &
 47 &   292    &   80 & 396 & 57 & 195 & 12 & 194 &
 15 \\
   604    &  946   &   7         &  428    &
 16 &   197    &   18 & 251 & 16 & 124 & 7 & 145
 & 9    \\
   605    &  948     &   6        &  351    &
 15 &   154    &   14 & 162 & 12 & 99 & 6 &
 86 & 7   \\
   606    &  942   &   7        &  359    &
 18 &   137    &   11 & 162 & 10 & 77 & 7 &
 77 & 8  \\
   607    &  954   &   7        &  304    &
 18 &   88    &   8 & 99 & 7 & 62 & 5 &
 64 & 6   \\
   608    &  935   &   9        &  289    &
 27  &   57    &   8 & 80 & 8 & 35 & 5 &
 48 & 6  \\
   609    &  923   &   14         &    325    &
 28 &   57    &   8 & 61 & 7 & 29 & 5 &
 10 & 5    \\
   610    &  929   &   16         &  344    & 
31 &   36    &      6 & 46 & 6 & 21 & 4 &
 28 & 5  \\ \hline
 592-596  &  1130    &   11         &  453    &
   27 &   72    &   6 & 76 & 5 & 39 & 4 & 32 & 4 \\
   597    &  1094    &   12        &  450    &
 32  &   145    &   13 & 140 & 20 & 54 & 8 &
 65 & 9   \\
   598    &  1075    &   6         &  348    &
 17 &   154    &   11 & 183 & 9 & 56 & 7 &
 88 & 8  \\
   599    &  1062    &   9         &  409    &
 23 &   161    &   24 & 231 & 20 & 98 & 8 &
 125 & 10   \\
   600    &  1050    &   10         &  622    &
 27 &   469    &   29 & 476 &  28 & 134 & 11 &
 192 & 13    \\
   601    &  1001    &   11         &  730    &
 35 &   580      &   32 & 657 & 34 & 149 & 12 &
 222 & 15   \\
   602    &  914   &   11         &  718    &
 33 &   633    &   35 & 629 & 34 & 144 & 13 &
 226 & 16   \\
   603    &  888   &   9         &  481    &
 24 &   373    &   36 & 328 & 28 & 123 & 9 & 116 &
 10 \\
   604    &  885   &   8        &  406    &
 21  &   196    &   19 & 214 & 17 & 108 & 7 & 
77 & 8  \\
   605    &  879   &   10         &  431    &
 25  &   160      &   17 & 170 & 16 & 87 & 8 &
 62 & 9   \\
 606-610  &  867   &   8          &  458    & 
17 &   82    &   5 & 92 & 5 & 45 & 3 & 46 & 4 \\ \hline 
 592-595  &  1099    &    22         &  515    &
 45 &   38    &   6  & 58 & 5 & 12 & 4 & 15 & 5    \\
 596-598  &  1101    &   17         &    512    &
 37 &   86    &   9 & 94 & 7 & 34 & 6 & 45 & 6   \\
 599-600  &  1090    &   14         &  489    &
 35 &   106   &   11 & 134 & 9 & 33 & 6 & 60 & 8    \\
 601-602  &  1077    &   11         &  498    &
 24 &   123    &   10 & 166 & 9 & 48 & 6 & 63 & 7     \\
 605-606  &  994   &    9       &   360   &
 23 &    78   &     8 &  110 & 7 & 38 & 5 & 48 & 6  \\
 607-610  &  980   &    11       &   372   &  
26 &     46   &    5 & 53 & 5 & 23 & 3 & 31 & 4    \\ \hline
 592-594  &  1095    &   11         &  342    &
 27 &   49    &   6 & 50 & 4 & 24 & 3 & 28 & 4   \\
 595-596  &  1100    &   15         &  468    &
 30 &   81    &    9 & 85 & 7 & 38 & 5 & 39 & 6    \\
 597-598  &  1079    &    18         &  519    &
 41 &   85    &   10 & 95 & 8 & 35 & 6 & 24 & 7    \\
 599-600  &  1018    &   19         &  549    &
 37 &   96    &   11 & 115 & 9 & 38 & 6 & 41 & 7      \\
 601-602  &  1017    &   20         &    541    &
 37 &   90    &   10 & 97 & 8 & 30 & 5 & 37 & 7   \\
 603-605  &   921    &   21        &  548    &
 41  &   86    &   9 & 81 & 7 & 29 & 5 & 27 & 6    \\
 606-610  &  811   &   15         &    383    & 
33 &   34    &   5 & 42 & 4 & 16 & 3 & 18 & 3    \\ \hline  
\end{tabular}
\end{table*}
\end{appendix}

\end{document}